# SIMULATION PLATFORM FOR MULTI AGENT BASED MANUFACTURING CONTROL SYSTEM BASED ON THE HYBRID AGENT

Ali Vatankhah barenji[1], Amir Shaygan[2], Reza Vatankhah Barenji[2]

[1]Department of Mechanical Engineering Eastern Mediterranean University, Famagusta, North Cyprus, Turkey

[2]Department of Industrial Engineering Hacettepe University, Beytepe Campus Ankara, Turkey

Ali.vatankhah@cc.emu.edu.tr

**ABSTRACT**

Agent based distributed manufacturing control and scheduling systems are subsets of new manufacturing systems. Multi agent systems (MAS) not only drive design and engineering control solutions but also influence flexibility, agility, and re-configurability, which makes MASs a better centralized systems than its traditional counterparts. However, implementation of all MASs in the real factories are timely, also extremely costly. A simulation environment that would allow independent development and testing of the services and business processes of the related manufacturing hardware is needed. This paper presents the design and implementation of a user-friendly simulation platform for multi agent based manufacturing control systems by considering the shop floor level. The proposed simulation platform can simulate the software level of the factory by considering the hardware level of the mentioned factory. An example of the simulation platform is presented for a flexible manufacturing system, which is located in EMU CIM lab.

## 1. INTRODUCTION

The era of "agile manufacturing", "lean manufacturing" and "Intelligent manufacturing" in recent years, has brought worldwide competition among the manufacturing enterprises. This competition between the stockholders in global market intensively returns to scheduling and control system performance of the production system. Recent developments in MASs in manufacturing control field can solve the existing problems. In order for the aforementioned problems to be solved, more flexibility and agility are required as mentioned by Marik et al. [1]. MASs offers an alternate approach to design and managing of the control systems by providing modularity, robustness and autonomy. Jennings et al. [2] has indicated that at least 25% of the manufacturing problems can be solved.

Jennings [2] also mentions that the software required for centralized methods is more complex than of those needed by the agents based approaches in terms of facilitating development, debugging and maintenance. However, the main problems of MASs are analyzing, testing, and validating of the behavior of the agent-based systems. Analyzing the MASs in manufacturing control not only depends on the software level but also on the hardware level in the shop floor. The design phase which takes place before the deployment of the resulted design in the real operation is usually an arduous and time consuming task. Therefore, simulation tools are required in order to support the correction of misunderstandings and errors before implementation as mentioned by Luke et al.[3].

Several environments are reported for the simulation of multi agent systems in the literature such as Vrba [4-6]. Barbosa et al. [7] used an agent based modeling platform for the simulation of multi agent manufacturing systems .However; these platforms are developed for specific cases and according to the application particularities. This means that simulating the behavior of agent based manufacturing systems for newly developed systems needs significant effort. In addition, there is a lack of standard platform with the ability to analyze multi agent based manufacturing control system in both hardware and software levels.  This paper proposes an architecture for simulating multi agent based manufacturing control system in the shop floor based on Hybrid Agent (HA). The HA is in simultaneous communication with MASs and shop floor control system. The remainder of this paper is structured as follows: Section 2 describes the architecture of simulation platform. Section 3 explains existing multi agent system. Section 4 explains a case study and discusses the results. Finally, Section 5 concludes the paper and presents future work.

## 2. ARCHITECTURE OF THE SIMULATION PLATFORM

Virtual reality is one of the ways to simulate manufacturing control systems by considering shop floor. However due to the extended development time associated with the virtual resources of each machine, analyzing this type of system in an actual factory setting is infeasible according to Lin et al.[8]. Additionally, based on Barenji [9],when comparing alternate scheduling systems, it is extremely intricate to replicate a similar condition for tentative tests .A prototype simulation that acts as an actual system can be used to overcome this problem. Moreover, it is indistinctly linked to the simulation platform or the actual system to control the actual system. In this research, an integrated simulation system i.e., "Simulation Platform" has been developed, following the approach of AV Barenji et al. [10] for testing MASs. The architecture of simulation platform is shown in Fig1.

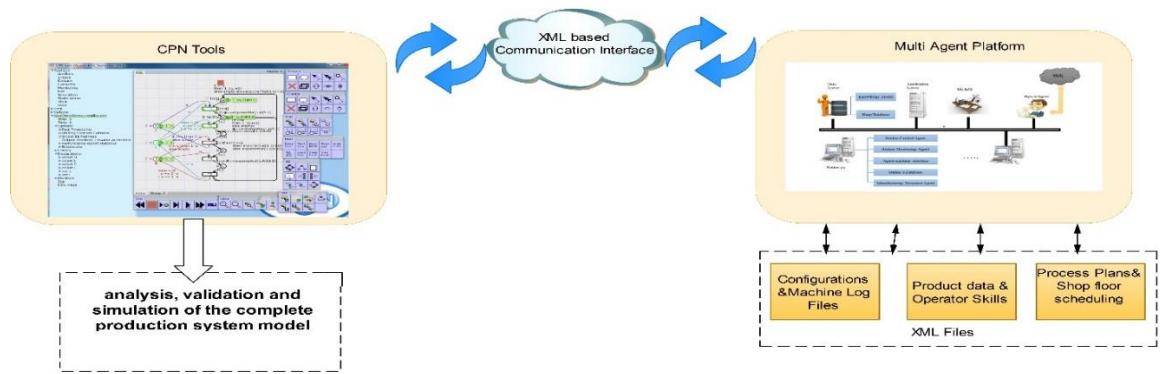

**Figure 1 Architecture of Simulation Platform**

The Simulation Platform consist of three main parts:

1. The Hardware Simulation Agent (HSA): HSA is a software which can design the hardware level of the factory by considering characterized physical actions that take place in the manufacturing environment. Examples of HSA are Arena, CPN tool, etc.
2. Multi agent based manufacturing control system: This part is based on the "Manufacturing Execution System" (MES) which is responsible for operation scheduling, production dispatching, work in process status, data acquisition, etc.
3. Hybrid agent: HAs are computational processes capable of moving throughout a network, interacting with foreign hosts, gathering information on behalf of the user, and returning to the user after performing their assignments. Characteristic of the HA are : a) Hybrid agent is capable of real time communication with other agent, b) HA is able to understand the decision taken by MASs, c) Hybrid agent is capable of communicating with HSA via XML (Extensible Markup Language).

The Multi agent based manufacturing control system are used to implement different kinds of MAS architecture with hybrid agents. Furthermore, the HSA characterizes the physical actions that take place in the manufacturing environment. This part is supported by many kinds of software such as Arena, PN tools, Flexisim and etc. The most characteristics of HSA software just ability to communication with XML. The primary feature of a simulation platform is to support the examination of any control system when applied to a discrete manufacturing system. Colored Petri Nets (CPN) which is a mathematical and graphical language for the design, specification, simulation, and verification of discrete systems can be named as an example of hardware simulation platform (HSP) software based on Tsinarakis et al. [11]. CPN and their extensions are widely used in modeling discrete-event dynamic systems, including production systems and networks. One of these tools that is used broadly is the CPN Tools which is a simulation software for the altering, simulating and analyzing Colored Petri Nets. This application can communicate via XML.

### 3. MULTI-AGENT SYSTEMS

An RFID (Radio Frequency Identification System) based multi-agent scheduling and control architecture has been developed in authors previous works named RFIDMASs in Barenji et al. [12-15].It should be mentioned that RFID are systems that transfer data with the aid of electromagnetic fields in order to automatically trace and identify the tagged entities. The proposed multi-agent system is designed as a network of software agents that interact with each other and with the system actors. These agents are categorized as shop management agents, agent managers, shop monitoring and command agents, station control agents, station monitoring agents, agent machine interfaces, and manufacturing resource agents. In addition to the agents forming the architecture, two databases exist in the architecture; a shop database and a station database. Furthermore, based on Barenji et al.[16], ontology (a capability-based knowledge model) is required for proper communication between the agents.

To illustrate the functionality of the proposed simulation platform for performance measurement of RFIDMASs, the Flexible Manufacturing Systems Laboratory (FMS Lab.) located in the Eastern Mediterranean University (EMU) is used as the case study.

## 4. CASE STUDY

The FMS laboratory of EMU consists of three stations: Station 1 is a machine tending station, which consists of a CNC milling machine and a five-axis vertically articulated robot. Station 2 is an assembly and quality control station, gluing machine, and laser-scan micrometer device. Station 3 is an automatic storage and retrieval system (AS/RS). A conveyer integrates the stations for performing material handling within the cell. RFIDMAS is selected as manufacturing execution system and CPN Tools version 4 is chosen as HSA. Fig2 shows the CPN model of the FMS lab.

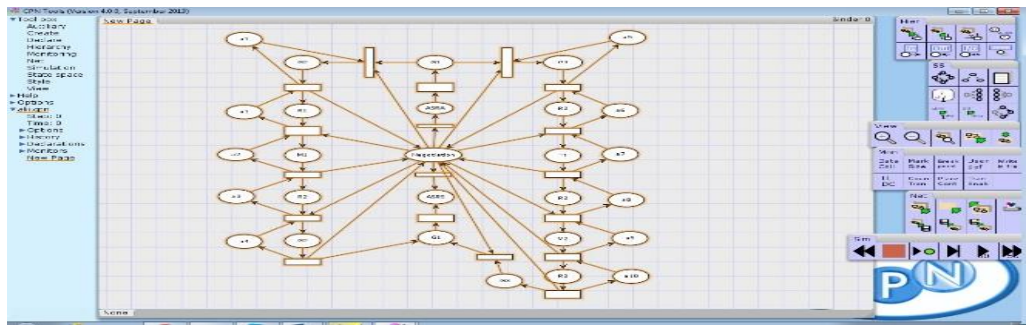

**Figure 2 CPN Model of the FMS Lab**

### 4.1 Hybrid agents

As mentioned in the part 2, hybrid agents are computational processes which are capable of maneuvering in a network, communicating with alien hosts, collecting data vicariously for the user, and returning to the user after performing its duty. Furthermore, hybrid agents have the ability of administrating the information existing in networks. Some of hybrid agents' features and capabilities are the real time inter-agent communication as well as their capability to comprehend the decision making process done by RFIDMASs. Additionally they possess the ability to exchange information with other software using XML. Fig 3 shows the sequence diagram for starting a new task in the system and illustrates the interaction between agents.

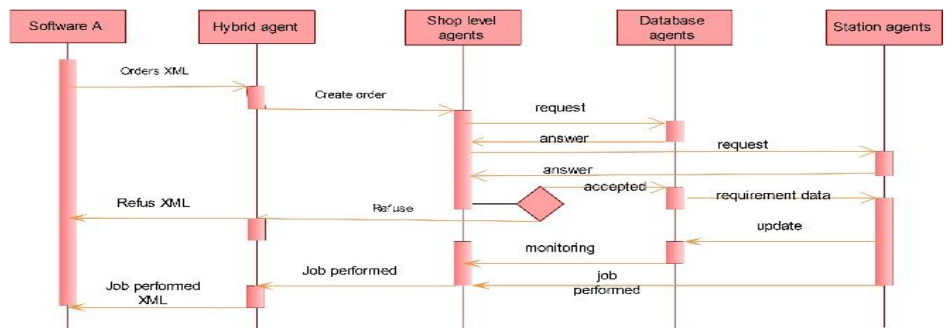

**Figure 3 Sequence Diagram for Starting a New Task in the System**

New task is sent to the HA via an XML file in order to communicate with shop-level agents. Shop-level agents consist of three types of agents, namely SMA, SMCA, and AM. A shop-level agent sends a request to a database agent for obtaining new data, and the database agent responds to this request. If the answer is positive, shop-level agents send a further request to the station agents checking for availability and possibility of doing a new job. If the answer to this request is positive as well, the shop-level agent accepts this job, sends a comment to the database agents, and the database agents send requirement data to the station agents. The station agent performs its duty via communication between sub-agents.

The hardware simulation platform (HSP), as mentioned before, defines the tangible actions taking place in the manufacturing environment. The communication model is set up in CPN Tools software, which creates the real time communication between the simulation part and multi agent control part. The XML codes for description of agents, MASs and resources are shown in the Fig4.

```
MAS:                                    Agent:                              Resources
    <MAS NAME= "RFIDMAS">                   <AGENT NAME= "SMG">                 <OBJECTS-LIST>
        <AGENTS-LIST>                           <ATTRIBUTES>                        <OBJECT NAME="ASRA">
            <AGENT NAME="HA"? >                     .                                   <ATTRIBUTES>
            </AGENT>                                .                                       .
                .                                   .                                       .
                .                               </ATTRIBUTES>                           </ATTRIBUTES>
        <OBJECT-LIST>                           <CURRENT-STATE>                         <CURRENT-STATE>
            <OBJECTNAME="ASRS">                     .                                       .
            </OBJECT>                               .                                       .
                .                               </CURRENT-STATE>                        </CURRENT-STATE>
                .                               <ACTIONS>                           </OBJECT>
        <STATES-LIST>                               .                               <OBJECTS-LIST
            .                                       .
            .                                   </ACTIONS>
        <ACTIONS-LIST>                      </AGENT>
    </MAS>
```

**Figure 2 XML Codes for Description of Agents, MASs and Resources**

The experiments is carried out by considering two different scenarios: (A) no disturbances and a well-functioning system and (B) incidents of failures in station one e.g. CNC, with a probability of 20%. In the experiments, setup time is not considered and it is assumed that negotiation failure would never occur. A robot and conveyor perform the transportation operations and orders are queued and executed in the arrival of the order. Each action of transportation takes 8 seconds. The process time for the CNC for each part is 10 seconds. Additionally, the process time of the assemble machine for the final product is 15 seconds. Each individual book of orders involves the production of 3 parts: 1 bodies, 1 handle, and 1 cover. The simulation-based reported test reflects 1000 book orders.

Performance indicators:

- Manufacturing lead time: The total time required to manufacture an item, including order preparation time, queue time, setup time, process time, move time, inspection time, and put-away time.

- Throughput: An indicator of the productivity of a manufacturing system, defined here as the number of items produced per time unit.

- Repeatability: The mean value of the standard deviation of the percentage of utilization of all resources of the system over several runs.

## 5. RESULTS AND DISCUSSION

The results from the simulation platform allowed us to draw some conclusions concerning the operation of the RFIDMASs in the FMS by taking into account the hardware and softer level. The system was found to function robustly and as specified in both normal operations and in the presence of disturbances. Furthermore, the re-configurability of the system is demonstrated by its accurate reactions to the introduction, removal, and modification of manufacturing components. In particular, it is shown that when a resource control agent broke down or was removed from the system, other agents continued to find alternative solutions for executing the production plan. Fig 5 shows the results for the stable scenario, without the occurrence of unexpected disturbances. In the stable scenario, the RFIDMASs system for the FMS yielded smaller values for the manufacturing lead-time (198) and higher values for the throughput (68), in comparison to those obtained using a conventional control system. The better performance of the proposed system is resulted by the cooperation of autonomous entities, i.e., an agent manger that elaborates optimized production plans.

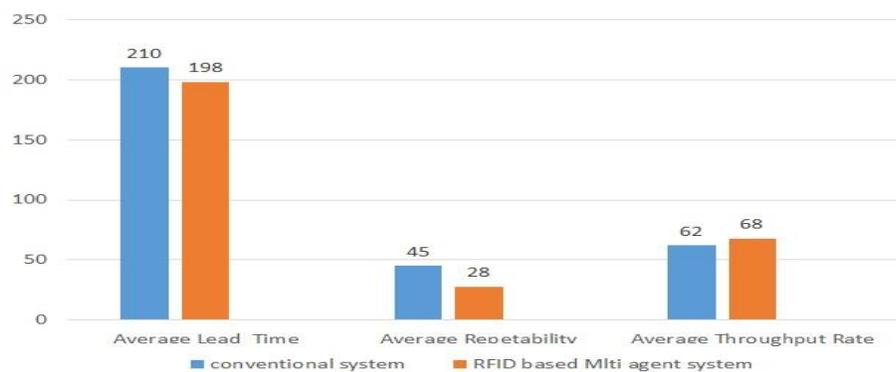

Figure 3  Stable Scenario

The results for the second scenario is summarized in Fig6. The first conclusion drawn from these simulation results is that the values of all performance indicators decreased in the presence of disturbances. An analysis of the lead times and throughputs confirmed that the RFIDMASs nonetheless yielded better performance than the conventional control systems.

Disturbances increase the entropy and unpredictability of a manufacturing control system. The implementation of a multi-agent scheduling approach improves system performance by improving the system's ability to respond to disturbances, as indicated by the smaller values of the manufacturing lead-time and higher values of the throughput than those obtained using the conventional scheduling control approach. The results indicate that the proposed system can achieve good productivity even when resource interruptions increase and that it can respond to resource breakdowns effectively. Analysis of the experimental results confirms that the use of the proposed multi-agent scheduling approach results in better resource utilization than the conventional approach in both stable and unstable scenarios.

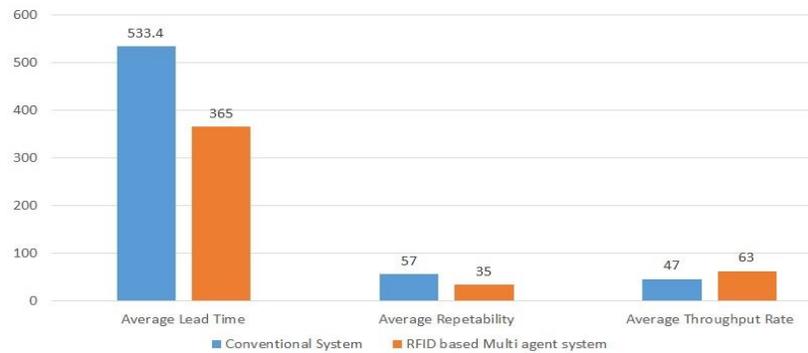

Figure 4  Disturbances Scenario

6. CONCLUSION

This paper proposes simulation platform for MASs based on the hybrid agent communication via XML coded with other software. A simulation-based performance assessment is performed to investigate the effectiveness of the multi-agent scheduling approach in comparison to a conventional scheduling approach. A case study of a flexible assembly system in a medium-sized factory was conducted. The multi-agent scheduling and production control system was tested over an integrated experimental test bed developed based on a simulation model of the real factory, integrated with an external multi-agent-based control system platform. The results of the simulation platform for the case study show that MASs for FMS create more re-configurability and agility. Future work will include the development of more user-friendly engineering tools for modeling and simulation of multi-agent manufacturing control system software tools and Petri net engineering platforms and achievement of a smooth migration from virtual scenarios to real systems.